\documentclass[11pt,tightenlines,eqsecnum,floats,aps,nofootinbib,prd]{revtex4-2}
\usepackage{hyperref}
\usepackage{amsmath,amssymb,amsfonts,amsthm,amscd}
\usepackage{enumerate}
\usepackage{colordvi}
\usepackage{units}
\usepackage{epsfig}
\usepackage{natbib}
\usepackage{enumerate}
\usepackage{afterpage}
\usepackage[utf8]{inputenc}
\def\ba{\begin{eqnarray}}
\def\ea{\end{eqnarray}}
\def\be{\begin{equation}}
\def\ee{\end{equation}}

\newcommand{\pv}[1]{\frac{\partial}{\partial #1}}
\def\Lie{\mathfrak{L}}
\def\L{\mathcal{L}}
\def\H{\mathcal{H}}
\def\D{\mathbf{D}}

\begin{document}

\title{Herglotz Action for Homogeneous Cosmologies}
\author{David Sloan}
\email{d.sloan@lancaster.ac.uk}
\affiliation{Department of Physics, Lancaster University, Lancaster UK}

\begin{abstract}
\noindent We present an action from which the dynamics of homogeneous cosmologies can be derived. The action has no dependence on scale within the system and hence is more parsimonious in its description than the Einstein-Hilbert action. The form of the action follows that pioneered by Herglotz and hence allows for a direct interpretation of the system as being both autonomous and frictional.  

\end{abstract}

\maketitle

\section{Introduction}
\label{Sec:Introduction}

General Relativity describes the dynamics of geometry. In cosmology we work with a symmetry reduced theory in which we consider homogeneous, and usually also isotropic, spatial manifolds that evolve in time. These can be described in terms of their (fixed) topology and the dynamical variables which describe their size (the isotropic mode), and shape (the anisotropies). In the isotropic case the shape is fixed and we typically describe these solutions, the Friedmann-Lem\^aitre-Robertson-Walker models, in terms of a scale factor $a$ which determines their size. It is well-known that it is only relational changes in $a$ that are physically relevant, and we choose to fix this once per solution to a given value. 

Despite the common approach being to fix the choice of $a$ as a given time, there is another possibility: eliminate scale from the theory entirely. We are motivated to choose to eliminate scale following the success of gauge-independent variables in physics. In a previous work \cite{Sloan:2020taf} I have shown how the freedom to choose $a$ can be exploited to find an action formulation which is entirely independent of scale. The new action, following Herglotz's principle, makes precise the relationship between expansion in the models which retain a sense of scale and friction in those for which this has been excised. The question then arises as to whether this formulation can be extended to include not only isotropic models, but also those which have anisotropic (shape) degrees of freedom. In this work, we will see that this is indeed the case. 

The purpose of this paper is to provide an action formulation of cosmology which is compatible with the relationalist/Shape Dynamics viewpoint\cite{Shapes1,Shapes2,Shapes3,Shapes4,Koslowski:2021aga}, from which the evolution of shape can be deduced without ever making reference to scale. This approach has met with great success in treating the initial singularity of cosmology \cite{Through,Mercati:2019cbn,Sloan:2019wrz}, and by parallel work the central singularity of Schwarzschild black holes  \cite{Mercati:2021zmv} (and see \cite{Bianchi:2018mml,DAmbrosio:2018wgv} for interesting potential consequences), as well as a number of other interesting physical systems \cite{Shyam:2021ppn,Mercati:2021azv}. The reduced descriptions appear frictional \cite{DynSim,Sloan:2021hwx} which has implications for the interpretation of fundamental issues in physics and cosmology such as the existence of a gravitational arrow of time, and the status of the past hypothesis \cite{Barbour2014,Barbour2015,Gryb:2020wat,Gryb:2021qix}. The frictional nature of these descriptions also provide an explanation for the evolution of physical (Liouville) measures on phase space \cite{Gibbons:1986xk,Measure,Measure2,MinCoup}, and see \cite{Bravetti:2020tau} for a mathematical description. Very recently, a complete mathematical description of the contact reduction has been shown for both the Lagrangian and Hamiltonian formalisms \cite{BravettiJackmanMe}.

The space-times we will consider in this paper are the class A Bianchi models - those homogeneous (but not necessarily isotropic) space-times which admit a Hamiltonian description. Bianchi models are interesting for two reasons. The first is that they provide a test-bed for understanding gravitational singularities that are more complex than those admitted by the purely isotropic FLRW models. The second is that a conjecture due to Belinskii, Khalatnikov and Lifschitz posits that in the neighbourhood of a general singularity, the partial differential equations describing the complete behaviour are well approximated by ordinary differential equations which model a homogeneous cosmology at each space-time point \cite{BKL,Kamenshchik:2017ous,AHS}. Whilst unproven, there is significant evidence both numerical and analytic to support this conjecture \cite{AR,Berger,Ringstrom}. 

In previous work, the procedure was to begin with Einstein's equations and show that the resultant equations of motion could be formulated in such a way that the shape degrees of freedom never made reference to the scale, using the geometric system as a Wittgenstein's ladder to access the underlying relational dynamics. Here we show that this reduction can be made at an earlier stage; in reducing the action we can see that the geometric picture is equivalent to the symplectification of a contact (Herglotz) system. This explains why scale can be removed in the first place, and further gives some hint towards an underlying relational replacement for general relativity. 

This paper is laid out as follows: In the following section, section (\ref{Sec:Setup}), we introduce the geometric set-up of the space-time that we will consider, the Bianchi class A cosmologies, and show how these have a Lagrangian description which separates shape from scale. Then in section (\ref{Sec:ScalingSymmetries}) we show the general construction of scaling symmetries on configuration space in such Lagrangian theories, and the reduction that can be made to arrive at a Herglotz Lagrangian description. In section (\ref{Sec:Applications}) we show how these reductions work, first by demonstrating their use on a known system (the Kepler problem) then establish the effect of reducing the Einstein Hilbert action. This is followed by a generalization of the result to include a broad range of matter fields and choices of scaling. Finally in section (\ref{Sec:Discussion}) we give an overview of the results and some notes on the potential implications of our work and future directions.

\section{Geometrical Setup}
\label{Sec:Setup}

We will consider four dimensional space-times which are homogeneous but anisotropic. We take our space-time manifold $\mathcal{M}=I\times \Sigma$ wherein time ranges over $I$, an interval in $\mathbb{R}$ and the spatial slice $\Sigma$ is a homogeneous three-dimensional Riemannian space. The space of possible three metrics on our spatial slice can be decomposed in terms of the commutativity of the three Killing vectors $\xi_i$: 
\be [\xi_i,\xi_j] = C^k_{ij} \xi_k \ee
The structure constants, $C^k_{ij}$ can be decomposed into their components:
\be C^k_{ij} = \epsilon_{ijl}n^{kl} + a_i \delta^k_j - a_j \delta^k_i \ee
$\epsilon$ is the alternating tensor. We can describe the diagonal tensor, $n^{ij}$, by its eigenvalues $n^i$. When non-zero, we chose fiducial lengths on the spatial slice such that $n_i = \pm 1$. The physically interesting models are the class A Bianchi models - see \cite{Ellis}, and for a nice introduction to these in a Hamiltonian setting see \cite{Jha:2022svf}. In these $a_i=0$, hence all the information about the spatial slice is contained in the $n_i$. Chosing a co-frame $\sigma$ compatible with the Killing vectors we obtain
\be \label{SigmaComm} d\sigma^1 = n_1 \sigma^2 \wedge \sigma^3 \quad d\sigma^2 = n_2 \sigma^3 \wedge \sigma^1 \quad d\sigma^3 = n_3 \sigma^1 \wedge \sigma^2 \ee

We will thus consider metrics of the form 
\be \label{metric} g = -dt^2 + v^\frac{2}{3} \left(e^{-\frac{x}{\sqrt{3}} + y} \sigma_1^2 + e^{-\frac{x}{\sqrt{3}} - y} \sigma_2^2 + e^{\frac{2x}{\sqrt{3}}} \sigma_3^2 \right) \ee
where $v,x,y$ are functions of $t$ which contain the dynamics of the system. These closely match the Misner variables, and have been used extensively in recent work to show the behaviour of scale-free systems at the big bang singularity \cite{Through,Mercati:2019cbn,Sloan:2019wrz}. By relaxing our conditions somewhat we can also include Kantowski-Sachs space-times; these have metrics the follow the form of equation (\ref{metric}) but the commutativity of Killing vectors varies. These are of particular interest as they can correspond also to the interiors of black holes. Useful properties of the metrics we will consider are given in table (\ref{BianchiClassification}).

\begin{table}[ht]
\begin{tabular}{|c|c|c|c|c|c|c|}
\hline
Type & $n_1$ & $n_2$ & $n_3$ & $\sigma^1$& $\sigma^2$& $\sigma^3$   \\
\hline
I & 0 & 0 & 0 & $dx$ & $dy$ & $dz$  \\
II & 1 & 0 & 0 & $dx$ & $dy$ & $xdy+dz$ \\
VI & 1 & -1 & 0 & $dx$ & $-\sinh x dy - \cosh x dz$ &$ \cosh x dy + \sinh x dz$ \\
VII & 1 & 1 & 0 & $dx$ & $-\sin x dy - \cos x dz$ &$ \cos x dy - \sin x dz$ \\
VIII & 1 & 1 & -1 & $dx - \cosh y dz$ & $ \sin x dy + \sinh y \cos x dz$&$ -\cos x dy+\sinh y \sin x dz$ \\
IX & 1 & 1 & 1 & $dx - \cos y dz$ & $ \sin x dy + \sin y \cos x dz$&$ -\cos x dy+\sin y \sin x dz$ \\
KS* & \multicolumn{3}{c|}{N/A} & $dx$ & $dy$ & $\sin y dz$\\
\hline
\end{tabular}
\caption{The Bianchi class A models in terms of the constants $n_i$, with a choice of co-frame satisfying equation \ref{SigmaComm}. The Kantowski-Sachs model is also included for completeness though it does not have the same symmetry structure.  }
\label{BianchiClassification}
\end{table}

The Ricci scalar on the spatial manifold can be expressed as a product of a scale term and a `shape potential' $V_s$ which depends only on the aniostropic parameters $x$ and $y$: ${}^3 R=v^{-\frac{2}{3}} V_s$. In the case of Bianchi models, the shape potential is:
\ba V_s &=& \frac{1}{2} e^{-\frac{2x}{\sqrt{3}}} h_-^2 - n_1 e^{\frac{x}{\sqrt{3}}} h_+ + \frac{n_1^2}{2} e^{\frac{4x}{\sqrt{3}}} \\
    h_\pm &=& n_2 e^{-y} \pm n_3 e^{y}  \ea
whereas for the Kantowski-Sachs model it is 
\be V_s = 2e^{-\frac{x}{\sqrt{3}}} \ee
The familiar FLRW space-times which are flat ($k=0$), open ($k=-1$) or closed ($k=1$) are subsets of the type $I$, $VIII$ and $IX$ models respectively, in which $x=y=0$. In such cases $V_s$ is a constant. 

For the metric given in equation (\ref{metric}) the Einstein-Hilbert Lagrangian $L_{\rm EH}$ becomes
\be \label{EHLag} L_{\rm{EH}} = v \left(-\frac{2\dot{v}^2}{3v^2} +\frac{\dot{x}^2}{2} + \frac{\dot{y}^2}{2} - \frac{V_s}{v^\frac{2}{3}} \right) \ee
Here it is important to note that we have symmetry reduced our theory at the level of the Lagrangian, from which we will derive the equations of motion compatible with these symmetries. Care has to be taken when doing this in general, as this procedure is not guaranteed to commute. In doing this we have to appeal to the ``principle of symmetric criticality" \cite{Palais:1979rca}. It has been shown by Torre and collaborators \cite{Fels:2001rv,Torre:2010xa} that the symmetry groups must satisfy specific conditions for this principle to hold. In the case of the class B Bianchi spacetimes the principle is not valid, due to the presence of boundary terms in the variational principle \cite{10.1093/mnras/142.2.129}. Fortunately, for the class A models that we care about, this is not the case, and the symmetry reduction of the Lagrangian leads to the correct extremal points. 

\section{Scaling Symmetries in Lagrangian Systems}
\label{Sec:ScalingSymmetries}

Our goal in this section is to introduce the idea of a {\em scaling symmetry} in a Lagrangian system. We will split our configuration space into `scale' (the variable which rescales under the symmetry) and `shape' parts (those which do not) and show that where such symmetries exist the Euler-Lagrange equations for the shape variables are equivalent to the Herglotz-Lagrange equations of a system written purely in shape terms. A simple example would be a two dimensional harmonic oscillator: The Lagrangian for this system is:
\be L = \frac{\dot{r}^2}{2} + \frac{r^2 \dot{\theta}^2}{2} - \frac{r^2}{2} \ee
And we see that multiplying $r$ by a constant $\alpha$ throughout simply serves to increase $L$ by $\alpha^2$. The scaling leaves $\theta$ invariant. We can find the equation of motion for $\theta$ from the usual Euler-Lagrange equation, or by first finding the Euler-Lagrange equation for $S=\frac{\dot{r}}{2r}$, and then minimizing $S$ subject to this equation of motion, i.e.
\be \dot{S} = -2S^2 + \frac{\dot{\theta}^2}{2}-\frac{1}{2} \ee
which gives rise to 
\be \ddot{\theta}+4S\dot{\theta}=0 \ee
which is the Euler-Lagrange equation for $\theta$. This is no coincidence, but rather a consequence of the existence of the scaling symmetry. The harmonic oscillator case is a simple one, as the scaling symmetry was independent of time. In the next two subsections we will first show how to construct scaling symmetries in general, and then show how these can be used to form an action principle following Herglotz's form from which the shape equations can be found. 

\subsection{Configuration Space Scaling Symmetries}

Let us consider a physical theory defined through an action principle. Physical trajectories are those curves $\gamma$ on the tangent bundle $TQ$, over a configuration space $Q$, which minimize the action $S$. The curves are parametrized over some interval $I \in \mathbb{R}$, usually taken to be time. The action is expressed in terms of a function $L(q,\dot{q})$ on $TQ$:
\be S = \int_\gamma L(q,\dot{q}) dt \ee
Curves $\gamma$ on $TQ$ are restricted to being the tangent lifts of curves $\overline{\gamma}$ on $Q$. That is to say that along $\gamma$,
\be \dot{q} = \frac{dq}{dt} \rightarrow \gamma = \overline{\gamma} \times \dot{\overline{\gamma}} \ee
A configuration space scaling symmetry (CSSS) $\overline{\D}$ is a vector field on $Q \times I$ such that:
\begin{itemize}
\item[i] for all curves $\overline{\gamma}:I \rightarrow Q$ whose tangent lifts minimize the action $S$, the tangent lifts of the curve $\overline{\D} \overline{\gamma}$ also minimize the action. 
\item[ii] $ \D L = \Lambda L $ 
\end{itemize}
where we will call $\Lambda \in \mathbb{R}$ the {\em degree} of the scaling symmetry, and $\D$ is the tangent lift of $\overline{\D}$. Hence we can consider the interpretation of the action of $\D$. The most direct interpretation is that the action of $\D$ is to move between solutions of the Euler-Lagrange equations with differing initial values of $q$. 

As an example, consider the Kepler problem. The map $\{r,\theta,t\} \rightarrow \{\lambda r, \theta, \lambda^\frac{3}{2} t\}$ transforms one solutions into another which has the same eccentricity and angular perihelion, but differing semi-major axis. 
Without loss of generality, we can choose coordinates on $Q$ such that (locally) each CSSS $\overline{\D}$ can be expressed
\be \label{CSSSForm} \overline{\D} = x \pv{x} + B t \pv{t} \ee
For a particular coordinate $x \in Q$ - in our Kepler example this is $r$. We will call as CSSS {\em isochronal} in $t$ if $B=0$. Such vector fields can be expressed purely through their action on $Q$. Compatibility of the tangent lift reveals the form of $\D$:
\be \D: TQ\times I \rightarrow TQ \times I \quad \pi (\D \gamma) = \overline{\D} \overline{\gamma} \ee
which in turn means that for $\overline{\D}$ of the form in equation (\ref{CSSSForm}) we can write:
\be \D = A x \pv{x} + (A-B) \dot{x} \pv{\dot{x}} - B \dot{q} \pv{\dot{q}} + B t \pv{t} \ee
It is clear that $\D$ (and $\overline{\D}$) are only defined up to a choice of normalization; if $\overline{\D}$ is a CSSS, then so is $\alpha \overline{\D}$ for $\alpha \in \mathbb{R} \setminus \{0\}$. To remove this redundancy we can impose the further condition that $\Lie_\D \mu_L=\mu_L$ where $\mu_L$ is the Lagrange one-form
\be \mu_L = \sum_{q \in Q} \frac{\partial L}{\partial \dot{q}} dq \ee
and we can choose coordinates such that $A=1$, though in practice it is often simpler to leave $A$ general. 

Let us make a further time reparametrization such that $d\tau = x^B dt$, and denote derivatives with respect to $\tau$ with a prime. Thus we arrive at a system from which 
\be \D = x \pv{x} + x'\pv{x'} \ee
Hence $\D$ is isochronal in $\tau$. 

Our goal now is to identify all such trajectories, and produce an action entirely in terms of the quantities that are invariant under $\D$. 

\subsection{Herglotz's Principle}

Herglotz's principle is an extension of the usual Lagrangian principle of least action to consider Lagrangians which are a function of both elements of the tangent space $TQ$ over a configuration space $Q$, and the action itself. These have proven to be of particular interest when modelling systems that include friction-like elements. The extension of the Lagrangian systems onto odd-dimensional manifolds (contact manifolds) means that we can consider physical systems that are non-conservative and autonomous. Here I will briefly outline some of the features of Herglotz-Lagrangian systems. For a complete overview see the comprehensive reviews in \cite{Leon,doi:10.1063/1.5096475,deLeon:2021nkb}. 

The Herglotz-Lagrange equations come from minimizing
\be S = \int L^H(q,\dot{q},S) dt \ee
over curves $\gamma: I\rightarrow Q$. In parallel to the usual Euler-Lagrange equations, minimizing this action by considering variations $\delta q$ we obtain
\be \label{HL} \frac{d}{dt} \left(\frac{\partial L^H}{\partial \dot{q}} \right) - \frac{\partial L^H}{\partial S} \frac{\partial L^H}{\partial \dot{q}} - \frac{\partial L^H}{\partial q} = 0 \ee
It is clear that when $L$ is independent of $S$ we recover the usual Euler-Lagrange systems. In general these systems are non-conservative; the contact Hamiltonian
\be H^c = \dot{q} \frac{\partial L^H}{\partial \dot{q}} - L^H \ee
is not a constant in time. A direct application of equation \ref{HL} reveals
\be \dot{H^c} = \frac{\partial L^H}{\partial S} H^c \ee
Further, it is a direct consequence of the equations of motion that if $L^H$ is independent of some coordinate, $q_i$, then the corresponding momentum $\pi^i = \frac{\partial L^H}{\partial \dot{q_i}}$ is not a constant, but rather a {\em dissipated quantity} \cite{Bravetti:2020jev}, such that for non-zero $H^c$, $\frac{\pi^i}{H^c}$ is constant (or, equivalently for any two such $\pi^i,\pi^j$, which are non-zero $\frac{\pi^i}{\pi^j}$ is constant).


Given a Herglotz Lagrangian $L^H$ on $TQ$, let us introduce $\rho$ such that $\dot{\rho}=-\frac{\partial L^H}{\partial S}$. Then $L= e^{\rho}(L^H + \dot{\rho} S)$ is a Lagrangian and it is clear that $\D=\pv{\rho}$ is a CSSS of degree 1. The equations of motion generated by $L$ and $L^H$ coincide for all $q$ in $Q$. To see this note that the from the Herglotz-Lagrange equations we find:
\be \frac{d}{dt} \left(\frac{\partial L}{\partial \dot{q}} \right) = e^\rho\left(\dot{\rho} \frac{\partial L^H}{\partial \dot{q}} + \frac{d}{dt}\left(\frac{\partial L^H}{\partial \dot{q}} \right) \right) = \frac{\partial L}{\partial q}                                                                                                                                                                                                                                                                                                                                                                                                                                                                                                                                                                                                                                                                                                                                                                                                                                  \ee
This construction allows us to show that given a Lagrangian $L$, with a CSSS $\D=x\pv{x} +\dot{x}\pv{\dot{x}}$ then 
\be L^H = \frac{\partial L}{\partial x} \ee
is a Herglotz Lagrangian which generates the same dynamics on  $\frac{TQ}{\D}$ is the same as the dynamics generated by $L$. First let us note that the Euler-Lagrange equation for $x$ indicates that $S=\frac{\partial L}{\partial \dot{x}}$ is an action for $L^H$. Hence we can write 
\be L = x L^H + \dot{x} S \ee
and identifying $\rho=\log x$ allows us to use the above result to show that the dynamics are equivalent. 

A parallel description of our scaling symmetry can be seen by considering that induced action $\D^*$ of $\D$ on the phase space $T^*Q$ in the Hamiltonian setting. In this set-up the symmetry was described as a dynamical similarity \cite{DynSim} with consequences discussed in \cite{Gryb:2021qix,Sloan:2021hwx}. From the definition and normalization of $\D$ we find that the Lie derivative of the Hamiltonian, $\H$ and symplectic structure $\omega$ are :
\be \Lie_{\D^*} \H = (1-B)\H \quad \Lie_{\D^*} \omega = \omega \ee
Hence we can see that the contact Hamiltonian we arrive at by reducing a symplectic Hamiltonian system by dynamical similarity is the same as that which we arrive at by beginning with a Lagrangian theory, reducing by the CSSS to find and Herglotz Lagrangian and then performing a Legendre transformation. 
\section{Applications}
\label{Sec:Applications}

\subsection{Warm Up: The Kepler Problem}

Before we tackle the case of homogeneous cosmologies, let us first consider the Kepler problem. Here we will work in polar coordinates such that our configuration space is $Q=S\times\mathbb{R}_+$, and consider rescaling the radial coordinate $r$. For brevity of exposition, we set all couplings and masses to unity. Hence we consider a Lagrangian and Lagrange one-form:
\ba \label{LagKepler} L_{\rm K}  &=& \frac{\dot{r}^2}{2} + \frac{r^2 \dot{\theta}^2}{2} + \frac{1}{r} \nonumber \\
	\mu_{\rm K} &=& \dot{r} dr + r^2 \dot{\theta} d\theta \ea 
We note that the symmetries of the Kepler problem such a rescaling is well-known; it is the basis of the third law. However, our purpose is to illustrate how this allows the reduction of the description to a Herglotz Lagrangian theory. We can obtain the form of our vector field either by direct application of the definition of a CSSS to the above, or by considering the third law directly:
\be \overline{\D} = 2 r \pv{r} + 3 t \pv{t} \rightarrow \D = 2r\pv{r} -\dot{r}\pv{\dot{r}} - 3\dot{\theta} \pv{\dot{\theta}} +3t \pv{t} \ee
In this, the action of $\D$ is to move between solutions that have the same eccentricity, but differing semi-major axis. The form of $\D$ is highly reminiscent of Kepler's Third Law; in moving between these solutions we retain the relationship between the orbital period and length of semi-major axis: $T^2 r^{-3}$ is unchanged under the action of $\D$. Given this relationship, we are motivated to make a change of variable and time parametrization to
\be x = \sqrt{r} \quad d\tau = x^{-3} dt \ee
in which $\D$ takes the required form and the Lagrangian one-form can be expressed:
\be \label{LagKeplerTau} \D = x \pv{x} + x'\pv{x'} \quad L dt = x\left(\frac{2x'^2}{x^2} +\frac{\theta'^2}{2} +1 \right) d\tau \ee
In this form the scaling is made quite apparent as the term between parentheses is invariant under the action of $\D$, but the pre-factor rescales. The one-form we obtain in $\tau$ from this Lagrangian is:
\be \mu_\tau = \frac{4x'dx}{x} + x\theta' d\theta = x^3 \mu_{\rm K} \ee

We now have our Lagrangian in the form required to make the reduction to the Herglotz Lagrange system. In doing so, we will be eliminating the radial direction from our system, but retaining the relative expansion $x'/x$ (equivalently $r'/r$):
\ba S &=& \frac{\iota_\D \mu}{x} =\frac{4x'}{x} \nonumber \\ 
L^H &=& \frac{\partial L}{\partial x} = -\frac{S^2}{8} + \frac{\theta'^2}{2} + 1 \ea
The presence of the $1$ in the Herglotz Lagrangian may appear unusual to readers who are more familiar with the usual Euler-Lagrange formulation, in which such terms would be eliminated as boundary contributions, as their only effect would be to shift the value of the action by a constant. In the Euler-Lagrange case, a shift of $S \rightarrow S+\alpha$ for $\alpha \in \mathbb{R}$ would not affect the motion. However, note that in the Herglotz Lagrangian $\dot{S}$ is a function of $S$. Hence these extra terms play an important role in the contact form of the system; shifting $S$ by a constant factor alters the dynamics. Therefore such terms cannot be eliminated from the Herglotz Lagrangian. We can trace the appearance of the $1$ to the coupling of $r^{-1}$ in equation (\ref{LagKepler}). 

Our current description is only valid for zero energy solutions. However, a simple trick allows us to extend our description to include the non-zero energy solutions. In the Hamiltonian formulation we can promote the total energy $E$ to a momentum on an extended phase space such that the Hamiltonian contains no conjugate position, and hence the energy is a constant. By choosing an appropriate power of this momentum we can ensure that the energy scales appropriately in the new time parametrization. This amounts to choosing the power such that the new term is invariant under the transformation $\D^*$ on phase space. The equivalent in the Lagrangian set-up is to introduce a new velocity such that the Lagrangian is independent of position. We see that if we supplement equation (\ref{LagKepler}) by adding $\dot{\chi}^n$ then in the Euler-Lagrange equations, $\dot{\chi}$ will be a constant in $t$. To retain the form of equation (\ref{LagKeplerTau}) we see that $n=\frac{2}{3}$ and hence we arrive at:
\be Ldt = x \left(\frac{2x'^2}{x^2} + \frac{\theta'^2}{2} +1 \pm \chi'^{\frac{2}{3}} \right) d\tau \ee
wherein the choice of $\pm$ represents positive or negative total energy. This choice can be eliminated by selecting an overall sign. However, since we are taking fractional powers of a $\chi'$ we would be required to consider $\chi' \in \mathbb{C}$, or more precisely $\chi'$ would be a product of a real number that changed over time and a (constant) root of unity. Whilst technically correct and formally providing solutions to the equations of motion, it is simpler to resolve to work with real $\chi'$ and specify the sign in the Lagrangian. In this case, we will choose the negative sign so that we consider closed orbits. 

Reducing our Lagrangian to the Herglotz Lagrangian following the above we find:
\ba L^H &=& -\frac{S^2}{8} + \frac{\theta'^2}{2} + 1 - \chi'^{\frac{2}{3}} \nonumber \\
    \H^c &=& \frac{S^2}{8} + \frac{\pi_\theta^2}{2} -1 + \frac{4}{27\pi_{\chi}^2} \ea
Note that $\pi_\chi$ is a dissipated quantity in the contact Hamiltonian formulation, and the powers thus serve to ensure that the rate of dissipation is in agreement with that we would have for the energy in the usual symplectic system when expressed in this time parametrization. 
with equations of motion
\ba \theta'' + \frac{S \theta'}{4} &=& 0 \nonumber \label{KepEOM} \\
    \chi'' -\frac{3S\chi'}{4} &=& 0 \ea
Which we can combine to show that $\theta' \chi'^{\frac{1}{3}}$ is a constant (equivalent to $J=r^2 \dot{\theta}$). From our definitions, it is simple to show that
\be \frac{\theta''}{\theta'} = -\frac{\dot{r} \sqrt{r}}{2} = -\frac{S}{4} \ee
which, together with the constant above can be used to show that both equations (\ref{KepEOM}) are results of symmetries - arising since the Lagrangian is independent of $\theta$ and $\chi$. The dynamics due to the potential are seen through the behaviour of $S$ which illustrates a common an important feature of these systems: We can infer apparent changes of scale in the Euler-Lagrange systems through the effect of the frictional terms in the scale free systems. If we model a potential which has a different fall-off in $r$ (e.g. $1/r^3$) we would see the difference manifest in the Herglotz Lagrangian by the coefficient of $S^2$ changing. 

\subsection{Homogeneous Cosmologies}

In the case of the Einstein-Hilbert Lagrangian given in equation (\ref{EHLag}) we can see that 
\be \overline{\D} = \frac{3}{2} v \pv{v} + \frac{1}{2} t\pv{t} \ee
is a CSSS of degree 1/2, with tangent lift
\be \D = \frac{3v}{2} \pv{v} + \dot{v} \pv{\dot{v}} - \frac{\dot{x}}{2}\pv{\dot{x}} - \frac{\dot{y}}{2} \pv{\dot{y}} + \frac{t}{2}\pv{t} \ee
we are therefore motivated to make the change of variables and time parametrization where $\rho=v^\frac{2}{3}$ and $d\tau = v^{-\frac{1}{3}} dt$. Under this transformation the Lagrangian is
\be L_{\rm{EH}} dt = \rho \left(-\frac{3\rho'^2}{2\rho^2} + \frac{x'^2}{2} + \frac{y'^2}{2} -V_s \right) d\tau \ee
and in this form $\D=\rho \pv{\rho} + \rho' \pv{\rho'}$. It is interesting to note at this point that this is the Einstein-Hilbert Lagrangian that we would have obtained had we begun by working in a co-frame of fixed determinant with $\sqrt{\rho}$ an overall factor, i.e. 
\ba {}^4 \mathbf{\sigma} &=& \sqrt{\rho} \left(d\tau,e^{-\frac{x}{\sqrt{12}} + \frac{y}{2}} \sigma_1 , e^{-\frac{x}{\sqrt{12}} - \frac{y}{2}} \sigma_2 ,e^{\frac{x}{\sqrt{3}}} \sigma_3 \right) \nonumber \\
ds^2 &=& \rho \left(-d\tau^2 + e^{-\frac{x}{\sqrt{3}} + y} \sigma_1^2 + e^{-\frac{x}{\sqrt{3}} - y} \sigma_2^2 + e^{\frac{2x}{\sqrt{3}}} \sigma_3^2 \right) \ea

 We are now in a position to find the Herglotz Lagrangian on $TQ/\D$ which has equivalent dynamics to $L_{\rm{EH}}$. Following the process given above, we find:
\ba S &=& \frac{\iota_{\D} \mu'}{\rho} = \frac{-3\rho'}{\rho} \nonumber \\ L^{H} &=& \frac{\partial L_{\rm{EH}}}{\partial \rho} = \frac{S^2}{6} + \frac{x'^2}{2} + \frac{y'^2}{2} - V_s \ea
from which we find the equations of motion
\ba x'' &=& \frac{Sx'}{3} - \frac{\partial V}{\partial x} \nonumber \\
    y'' &=& \frac{Sy'}{3} - \frac{\partial V}{\partial y} \nonumber \\
    S' &=& L^H = \frac{S^2}{6} + \frac{x'^2}{2} + \frac{y'^2}{2} - V_s. \ea
and hence
\be S = \int  \frac{S^2}{6} + \frac{x'^2}{2} + \frac{y'^2}{2} - V_s d\tau \label{MainAction} \ee
is a Herglotz Lagrangian which reproduces the equations of motion of the Einstein Hilbert Lagrangian for the shapes. 

We can recover the contact Hamiltonian, $\H^c$, from the Herglotz Lagrangian, $L^H$, using a Legendre transform:
\be \H^c = -\frac{S^2}{6} + \frac{\pi_x^2}{2} + \frac{\pi_y^2}{2} + V_s. \ee

Let us now extend the description of the system to include homogeneous matter. For clarity of exposition we will consider the role of matter to be played by a scalar field $\phi$ with potential $U$, but the construction is general. We begin with the Einstein-Hilbert Lagrangian, equation \ref{EHLag}, and add a minimally coupled homogeneous scalar field:
\be L dt = \L_{\rm EH} dt + \L_\phi dt=  v \left(-\frac{2\dot{v}^2}{3v^2} +\frac{\dot{x}^2}{2} + \frac{\dot{y}^2}{2} - \frac{V_s}{v^\frac{2}{3}} +\frac{\dot{\phi}^2}{2} - U(\phi) \right) dt \ee
In this for $\overline{\D}$ is no longer a CSSS, as the potential term, $vU$ does not rescale with degree 1. However, following the Kepler example, we can introduce a new field $\kappa$ such that 
\be L = v \left(-\frac{2\dot{v}^2}{3v^2} +\frac{\dot{x}^2}{2} + \frac{\dot{y}^2}{2} - \frac{V_s}{v^\frac{2}{3}} +\frac{\dot{\phi}^2}{2}\right) + \sqrt{v \dot{\kappa} U(\phi)  } \ee
we find that we reproduce the same equations of motion as our prior Lagrangian, as the Euler-Lagrange equations for $\kappa$ and $\phi$ yield:
\be \dot{\kappa} = C U v \quad \mathrm{and} \quad  v\ddot{\phi} + \dot{v}\dot{\phi} = \sqrt{\frac{v\dot{\kappa}}{4U}} \frac{\partial U}{\partial \phi} = \frac{vC}{2} \frac{\partial U}{\partial \phi} \ee
and imposing the boundary condition $C=2$ gives rise to the usual equations of motion. Further, $\overline{\D}$ is once again a CSSS for this Lagrangian. 

At this point it is worth a minor digression to note that one may be tempted to claim that the new Lagrangian is more complicated than the former, as we have introduced a field $\kappa$ and required a boundary condition to match the original dynamics. However, the boundary condition we have introduced is equivalent to a choice of an overall scale of the potential $U$. To make this explicit, consider the simplest potential $U=m^2 \phi^2$: in the original Lagrangian we would have to not only specify initial values for each of the fields, but also the value of $m$. In the reformulation we have simply traded specifying the value of $m$, a constant, for a boundary condition specifying the value of $C$. In terms of the total number of external inputs required, therefore, there is no cost in transforming the Lagrangian in this manner. 

Following the same changes of variable and time parametrization as above we arrive at 
\be L dt = \rho \left(-\frac{3\rho'^2}{2\rho^2} + \frac{x'^2}{2} + \frac{y'^2}{2} -V_s + \frac{\phi'^2}{2} + \sqrt{\kappa' U}) \right) d\tau \ee
and we can apply the same process to arrive at the Herglotz Lagrangian
\be L^H = \frac{S^2}{6} +\frac{x'^2}{2} + \frac{y'^2}{2} +\frac{\phi'^2}{2} -V_s + \sqrt{\kappa' U} \ee
from which the Herglotz-Lagrange equations give rise to the same equations of motion as the original system. Further, the role of $\kappa$ is elucidated by considering the contact Hamiltonian
\be \H^c = -\frac{S^2}{6} + \frac{\pi_x^2}{2} + \frac{\pi_y^2}{2} + \frac{\pi_\phi^2}{2} +V_s + \frac{U}{2 \pi_\kappa} \ee
where we see that since $\H^c$ is independent of $\kappa$, $\pi_\kappa$ is a dissipated quantity which plays the same role as changing volume in the symplectic formalism. 

\subsection{Generalizations}

To deal with terms that are not compatible with a given CSSS we have constructed new, equivalent Lagrangians with extra velocity terms. These extra terms, due to the freedom to choose their powers, allow us to ensure that our CSSS is still valid, and have a term that, due to the frictional effect on velocities in the Herglotz framework, evolve over time in the same way as the volume terms in the original Lagrangian. We can, therefore, adopt an alternate position wherein we first choose the CSSS and then modify the Lagrangian to fit it, without altering the evolution of the observable (shape) quantities. 

Let us parametrize the space of CSSS that act on $TQ \times \mathbb{R}$ by the rescaling that they enact upon the purely kinetic parts of the Einstein-Hilbert Lagrangian, $K$ where
\be K = v\left(\frac{2\dot{v}^2}{3v^2} + \frac{\dot{x}^2}{2} + \frac{\dot{y}^2}{2} \right) \ee
Note that $K$ is the Lagrangian of a Bianchi I model, since the shape potential there is zero. Our definition of $\D_\Lambda$ is that $\Lie_{\D_\Lambda} K = \Lambda K$. In such case $\D_\Lambda$ is given:
\be \D_\Lambda = (2-\Lambda) v\pv{v} + \dot{v}\pv{\dot{v}} +(\Lambda-1) \dot{x} \pv{\dot{x}} +(\Lambda-1) \dot{y} \pv{\dot{y}} + (1-\Lambda) t\pv{t} \ee
Following the methods of the above, we find the Herglotz Lagrangian $K^H$ is
\be K^H = -\frac{3}{8} \left(\frac{S}{2-\Lambda}\right)^2 +\frac{x'^2}{2} + \frac{y'^2}{2} \ee
where 
\be S=\frac{4(\Lambda-2)}{3} \frac{\dot{v}}{v^\frac{1}{2-\Lambda}} \ee
We will follow our above constructions in adding velocities with no conjugate momenta to the Lagrangian to model the effect of the other fields. In doing so, we extend $\D_\Lambda$ by adding $(\Lambda-1)\dot{q_i} \pv{\dot{q_i}}$ such that $\D_\Lambda$ acts the same way on all velocities. 

We can introduce a term that reproduces the effect of the shape potential by adding 
\be L_s = v^\frac{1}{5-4\Lambda} V_s^\frac{3}{5-4\Lambda} \dot{s}^\frac{4\Lambda-2}{4\Lambda-5} \ee
noting that $\Lie_{\D_\Lambda} L_s = \Lambda L_s$, where we make the natural extension to add a term $(\Lambda-1) \dot{s}\pv{\dot{s}}$ to $\D_\Lambda$. This results in a term in the Herglotz Lagrangian
\be L^H_s = {s'}^\frac{2-4\Lambda}{5-4\Lambda} V_s^\frac{3}{5-4\Lambda} \ee
We can note two things here - the first is that in the case $\Lambda=\frac{1}{2}$ we indeed reproduce the action of equation \ref{MainAction} above in which $s'$ is not present. Further, if we choose $\Lambda=1$ we reproduce the Herglotz Lagrangian of \cite{Sloan:2020taf}.

Let us now consider terms that we can add to $K$ which will reproduce the dynamics of various matter fields. In particular let us consider homogeneous, isotropic perfect fluids with equation of state $P=w\rho$. From the continuity equation, 
\be \dot{\rho} + 3H(\rho+P) \ee 
we know that $\rho \sim v^{-(1+w)}$. Hence we can reproduce the effects of this by introducing new terms, $L_w$ that have the correct powers such that the frictional effect of the Herglotz Lagrangian matches this. We further want these to satisfy $\Lie_\D L_w = \Lambda L_w$. These turn out to be 
\be L_w = \left(v^w\dot{u}^{\Lambda+(2-\Lambda)w}\right)^\frac{1}{\Lambda+(2-\Lambda)w-1} \ee
which in turn lead to terms in the Herglotz Lagrangian of 
\be L^H_w = {u'}^\frac{\Lambda+(2-\Lambda)w}{\Lambda+(2-\Lambda)w-1} \ee
Hence we can reconstruct all the usual fluid and curvature terms and anisotropies that contribute to the evolution of homogeneous space-times in the Herglotz Lagrangian framework, with a choice of actions
\be S_\Lambda = \int \left( K^H + L^H_s + \sum L^H_w \right) d\tau \ee
which have CSSS of degree $\Lambda$. 

There is a direct benefit to this analysis when considering the issue of the big bang singularity, as this is a point at which, in the space-time picture, $v \rightarrow 0$. In our case we have eliminated $v$ from our equations, and introduced some dissipative quantities which play the same role. By choosing $\Lambda$ such that there is no velocity associated with the dominant field contribution at the singularity we are able to examine an equivalent system to the Einstein-Hilbert system which evolves purely in shapes. These have been shown to admit continuations through the singularity, and thus this set of actions provides a framework for assessing a broad range of such cosmological models. 
 
\section{Discussion}
\label{Sec:Discussion}

The existence of a scale free description of cosmology provides support for the position that relational motion (the evolution of `shapes' rather than `sizes') is fundamental. Such a position would hold that the geometric description of relativity as a four-dimensional space-time manifold is in fact a useful fiction that can be used as a tool for performing calculations, but should not have fundamental ontological status. As such problems that occur in the geometric description of physics, such as the existence of singularities, may be a failure of the fiction rather than a breakdown of fundamental physics. In recent works it has been shown that although the geometric description of cosmologies cannot be continued beyond, for example, the big bang, the relational description finds no such impediment and has a well-defined, deterministic evolution. Away from this point the geometric description can be again employed as a calculational tool, and agrees everywhere with the relational description. In this sense, scale can be seen as ``superfluous structure" \cite{Ismael,Gryb:2021qix}. The excision of such structure is often the route to uncovering more fundamental theory; through the removal of preferred reference frames of Newton's theory, a more fundamental description can be found. Since the scale factor is never directly observable, and all other dynamical variables are reproduced by the Herglotz action of equation (\ref{HL}), the shape theory ``saves the phenomena" (in the language of van Fraassen) \cite{VanFraassenBas1980-VANTSI} of the Einstein-Hilbert action.

The fact that cosmology with scale can be seen to arise as the symplectification of a more parsimonious contact system has important implications for interpretations of cosmology. Contact systems are naturally frictional as the Liouville measures is not conserved under time evolution, nor is the total energy (Hamiltonian) unless this is zero. In relativity, the latter is zero as it is a constraint that arises as a result of time parametrization invariance of the underlying theory. However, the non-conservation of the Liouville measure gives rise to the possibility of describing cosmology as an open system, which has recently been posited being more fundamental than their closed counterparts \cite{Cuffaro:2021jpy}. On first inspection it may appear unusual to describe cosmology in an open systems language; the universe as a whole is often considered to be the only example of a truly closed system. However, it is important to note the manner in which the system is open, and those in which it is not: although the measure is not conserved, there is no coupling to an external system, and hence no exchange of energy or information with a separate entity. The equations of motion are autonomous - they need no reference to external entities beyond the usual specification of initial (boundary) conditions. In this sense systems described by Herglotz Lagrangians (equivalently contact Hamiltonians) are closed in terms of a dynamical algebra. 

The continuation of the classical shape equations of motion through a singularity is an enticing prospect. Although the space-time picture may not be well defined at this point, it does appear that a number of physical quantities can be extended through this point. Alongside the continuation of a sufficient number of shape quantities to determine a unique evolution, it has recently been shown that quantum fields \cite{Casadio:2020zmn,Ashtekar:2021dab,Ashtekar:2022oyq} (in the form of operator valued distributions) can also be extended beyond these points. This adds to a growing body of evidence that the initial singularity may not be the end of physics. This is potentially important for quantum gravity as existence and impassible nature of the initial singularity means it is one of the most frequently examined point in quantum approaches to gravity. If the singularity does not present such an impediment, it may be that quantum approaches that do not resolve the issue are not invalid.

Finally, let us note that much of the canonical approach to quantum cosmology is based on promoting $v$ (or some function thereof) to an operator, together with its conjugate momentum, and imposing commutativity relations. This is the basis of Wheeler-deWitt quantum cosmology, and lies at the heart of most of Loop Quantum Cosmology \cite{LQG2,LQC1}. In this and previous works, we have shown that this variable can be excised completely from the classical description. This points to an alternate quantization method, which should be based upon the contact geometry rather than its symplectic embedding. It is as yet unknown how to carry out this process. 

\section*{Acknowledgements}

The author is grateful to Alessandro Bravetti and Connor Jackman for helpful discussions and comments.

\bibliographystyle{ieeetr}
\bibliography{HerglotzCosmology2}

\end{document}